\documentclass{format170x240multiauthor}
\usepackage{graphics,pstricks,pst-node,epsf}
\usepackage{cite}
\usepackage[T1]{fontenc}
\usepackage{makeidx}
   \makeindex
\begin{document}
\chapter{Information Based Complexity of Networks}

\chapterauthor[Russell K. Standish]{Russell K. Standish\\
                Mathematics and Statistics, University of New South
                Wales\\
                hpcoder@hpcoders.com.au
        }

\section{Introduction}

Information is a measure of the amount of reduction in uncertainty the
receipt of a message causes in the receiver. It is also used
interchangably with the term complexity, referring to a measure of how
complex a system might be. 

Simple systems with few and regularly behaving parts require little
information to describe how the system behaves. Conversely, systems
with many similar parts may well admit a simple statistical
description that is also of low complexity. Random behaviour, in
particular, is of low complexity, as randomness, by definition,
entails that no specific model for the behaviour exists, and only
simple statistical descriptions are available.

By contrast, a system that needs to be modelled in great detail to
capture the essential behaviour, an automobile, or a living cell, is a
{\em complex system}, requiring a large amount of information to
specify the systems model.

Note that in the course of the preceding paragraphs, the terms {\em
models} and {\em descriptions} slipped in. Complexity (and indeed
information) is an observer dependent term~\cite{Standish01a}. What may
be simple for the intents and purposes of one observer may well be
complex to another. Nevertheless, once a discussion has been
adequately framed so that observers agree on what is important about a
system being discussed, information theory provides an objective
measure of the amount of information or complexity a system exhibits.

When talking about the complexity of networks, it is important to
realise that networks in themselves are abstract models of some
system. We need to be clear whether the nodes are distinguishable,
other than by their position within the network, by labels perhaps, or
categories such as colours. There may be dynamics between the parts of
the system represented by the network, which needs to be represented
in any consideration of complexity.

In what follows, starting with with unlabelled, undirected static
networks, we will consider the effects of labelling and colouring
nodes, directed edges between the nodes, weighted edges and finally
how to measure the complexity of a dynamical system defined on a network.

\section{History and concept of information based complexity}

%Information was initially quantified with the work of Claude Shannon,
%who introduced a measure of the amount of reduction in uncertainty the
%receipt of a message causes in the receiver. Given a message being
%transmitted with probability $p_i$, the information imparted to the
%receiver is $-\log_2 p_i$ bits. A source of such messages imparts an
%average amount of information 
%\begin{equation}
%H = -\sum_i p_i \log_2 p_i,
%\end{equation}
%a quantity Shannon called {\em entropy} by analogy with the well-known
%Gibbs formula for thermodynamic entropy.

Information theory began in the work of Shannon
~\cite{Shannon49}, who was concerned with the practical problem of
ensuring reliable transmission of messages. Every possible message has
a certain probability of occurring. The less likely a message is, the
more information it imparts to the listener of that message. The
precise relationship is given by a logarithm:
\begin{equation}\label{shannon}
I = -\log_2 p
\end{equation}
where $p$ is the probability of the message, and $I$ is the
information it contains for the listener. The base of the logarithm
determines what units information is measured in --- base 2 means the
information is expressed in {\em bits}. Base 256 could be used to
express the result in {\em bytes}, and is of course equivalent to
dividing equation (\ref{shannon}) by 8.

\begin{figure}
\fbox{
\psset{xunit=0.75cm,yunit=0.75cm}
\begin{pspicture}(0,1)(15,9)
%\psgrid
\rput[l](0,5){\epsfxsize=6cm\epsfbox{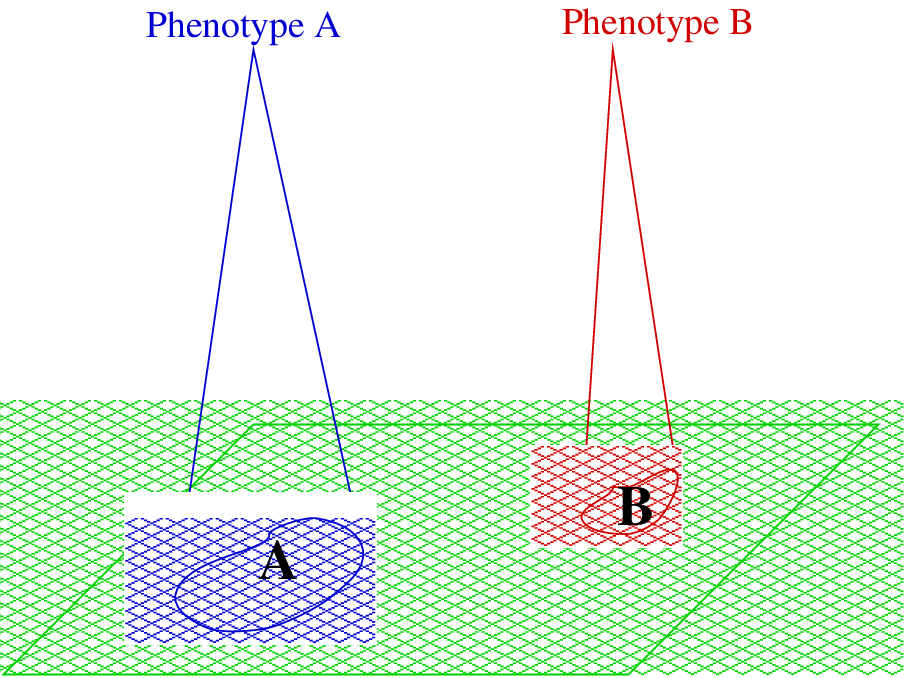}}
\rput(2.5,2.4){{\blue\psframebox*[framearc=.3]{${\cal A}$}}}
\rput(5.8,3){{\red\psframebox*[framearc=.3]{${\cal B}$}}}
\rput[l](8,2){${\cal L}_1$ Syntactic layer}
\rput[l](8,7){${\cal L}_2$ Semantic layer}
\rput[l](8,5){\parbox{5cm}{B is more complex (or has greater
    information) than A, because the set ${\cal B}$ is smaller
  than ${\cal A}$}}
\end{pspicture}
}
\caption{Diagram showing the syntactic and semantic spaces. Two
  different messages, having meanings A and B, can each be coded in
  many equivalent ways in syntactic space, represented by the sets
  ${\cal A}$ and ${\cal B}$. The information or complexity of the
  messages is related to the size it occupies in syntactic space by
  formula (\ref{shannon})}.
\label{syn-sem}
\end{figure}

Shannon, of course, was not so interested in the semantic content of
the message (ie its meaning), rather in the task of information
transmission so instead considered a message composed of symbols $x_i$
drawn from an alphabet $A$. Each symbol had a certain probability
$p(x_i)$ of appearing in a message --- consider how the letter `e' is
far more probable in English text than the letter `q'. These
probabilities can be easily measured by examining extant texts. A first
order approximation to equation (\ref{shannon}) is given by:
\begin{equation}\label{shannon entropy} % hey check this out!!!
I(x_1x_2\ldots x_n) \approx \sum_{i=1}^n p(x_i)\log_2 p(x_i)
\end{equation}
This equation can be refined by considering possible pairs of letters,
then possible triplets, in the limit converging on the minimum amount
of information required to be transmitted in order for the message to
be reconstructed in its original form. That this value may be
considerably less that just sending the original message in its
entirety is the basis of compression algorithms, such as those
employed by the well-known {\em gzip} or {\em PKzip} (aka WinZip)
programs.

The issue of semantic content discouraged a lot of people from
applying this formalism to complexity measures. The problem is that a
message written in English will mean something to a native English
speaker, but be total gibberish to someone brought up in the Amazon
jungle with no contact with the English speaking world. The
information content of the message depends on exactly who the listener
is! Whilst this context dependence appears to make the whole
enterprise hopeless, it is in fact a feature of all the naive
complexity measures normally discussed. When counting the number of
parts in a system, one must make a decision as to what exactly
constitutes a part, which is invariably somewhat subjective, and needs
to be decided by consensus or convention by the parties involved in
the discussion. Think of the problems in trying decide whether a group
of animals is one species of two, or which genus they belong to. The
same issue arises with the characterisation of the system by a
network. When is a relationship considered a graph edge, when often
every component is connected to every other part in varying degrees.

However, in many situations, there appears to be an obvious way of
partitioning the system, or categorising it. In such a case, where two
observers agree on the same way of interpreting a system, then they
can agree on the complexity that system has. If there is no agreement
on how to perform this categorisation, then complexity is meaningless

To formalise complexity then, assume as given a classifier system that
can categorise descriptions into equivalence classes. This is sketched
in Figure \ref{syn-sem}, where sets of descriptions in the syntactic
layer ${\cal L}_1$ are mapped to messages in the semantic layer ${\cal
L}_2$. Clearly, humans are very good at this --- they're able to
recognise patterns even in almost completely random data. Rorschach
plots are random ink plots that are interpreted by viewers as a
variety of meaningful images. However, a human classifier system is
not the only possibility. Another is the classification of programs
executed by a computer by what output they produce. Technically, in
these discussions, researchers use a {\em Universal Turing Machine}
(UTM), an abstract model of a computer.

Consider then the set of possible binary strings, which can fed into a
UTM $U$ as a program. Some of these programs cause $U$ to produce some
output then halt. Others will continue executing forever. In
principle, it is impossible to determine generally if a program will
halt or continue on indefinitely. This is the so called {\em halting
problem}. Now consider a program $p$ that causes the UTM to
output a specific string $s$ and then halt. Since the UTM halts after
a certain number of instructions executed (denoted $\ell(p)$) 
the same result is produced by feeding in any string starting with the same
$\ell(p)$ bits.  If the strings have equal chance of being chosen
({\em uniform measure}), then the proportion of strings starting with the
same initial $\ell(p)$ bits is $2^{-\ell(p)}$. This leads to the {\em
universal prior} distribution over descriptions $s$, also known as the
Solomonoff-Levin distribution:
\begin{equation}\label{universal prior}
P(s) = \sum_{\{p: U(p)=s\}}2^{-\ell(p)}
\end{equation}

The complexity (or information content) of the description is given by
equation (\ref{shannon}), or simply the logarithm of (\ref{universal
prior}). In the case of an arbitrary classifier system, the
complexity is given by the negative logarithm of the equivalence class size
\begin{equation}\label{complexity}
{\cal C}(x) = \lim_{s\rightarrow\infty} s\log_2 N - \log_2 \omega(s,x)
\end{equation}
where $N$ is the size of the alphabet used to encode the description
and $\omega(s,x)$ is the number of equivalent descriptions having
meaning $x$ of size $s$ or less~\cite{Standish01a}.

It turns out that the probability $P(s)$ in equation (\ref{universal
  prior}) is dominated by the shortest program~\cite[Thm
4.3.3]{Li-Vitanyi97}, namely
\begin{equation}
K(s)+\log_2 P(s) \leq C
\end{equation}
($\log_2 P(s) <0$ naturally) where $C$ is a constant independent of
the description $s$. $K(s)$ is the length of the shortest program $p$
that causes $U$ to output $s$, and is called the {\em Kolmogorov
  complexity} or {\em algorithmic complexity}.

An interesting difference between algorithmic complexity, and the
general complexity based on human observers can be seen by considering
the case of random strings. {\em Random}, as used in algorithmic
information theory, means that no shorter algorithm can be found to
produce a string than simply saying ``print \ldots'', where the
\ldots{} is a literal representation of the string. The algorithmic
complexity of a random string is high, at least as high as the length
of the string itself. However, a human observer simply sees a random
string as a jumble of letters, much the same as any other random
string. In this latter case, the equivalence class of random strings
is very large, close to $N^s$, so the perceived complexity
is small. Thus the human classifier defines an example of what
Gell-Mann calls {\em effective complexity}~\cite{Gell-Mann94}, namely a
complexity that has a high value for descriptions that are partially
compressible by complex schema, but low for random or obviously
regular systems.

A good introduction to information theoretical concepts for complex
systems studies can be found in~\cite{Adami98a}.

\section{Mutual Information}

When considering information {\em transfer}, it is useful to consider
the amount of information transferred in a message to be related to
the reduction in uncertainty the receiver has about the source on
receipt of the message. In order to quantify this, consider the sender
and receiver to be stochastic variables $X$ and $Y$, and form the
joint probability:
\begin{equation}
P(X=x_i \ \mathrm{and}\  Y=y_j) = p(x_i,y_j)
\end{equation}
We can then form the entropies 
\begin{eqnarray*}
H(X)&=&\sum_i P(X=x_i)\log P(X=x_i)\\
H(Y)&=&\sum_i P(Y=y_i)\log P(Y=y_i)
\end{eqnarray*}
 and the joint entropy 
\begin{displaymath}
H(X,Y)=\sum_{ij}
p(x_i,y_j)\log p(x_i,y_j).
\end{displaymath}
 If the
processes $X$ and $Y$ are independent of each other, we have 
\begin{displaymath}
p(x_i,y_j) = P(X=x_i)P(Y=y_j),
\end{displaymath}
so therefore
\begin{displaymath}
H(X,Y) = H(X)+H(Y)
\end{displaymath}
for independent processes. In general, however
\begin{displaymath}
H(X,Y) \leq H(X)+H(Y).
\end{displaymath}
The difference is known as {\em mutual information}:
\begin{equation}\label{mutual information}
I(X:Y) = H(X)+H(Y)-H(X,Y)
\end{equation}

{\em Conditional entropy} is the usual entropy applied to conditional
probability $P(X=x_i|y_j)$:
\begin{equation}
H(X|Y) = \sum_{ij} p(x_i,y_j) \log P(X=x_i|y_j).
\end{equation}
Using Bayes rule, mutual information can be expressed in terms of te
conditional entropy as
\begin{equation}
I(X:Y) = H(X)-H(X|Y) = H(Y)-H(Y|X).
\end{equation}

\section{Graph theory, and graph theoretic measures: cyclomatic number,
     spanning trees}

Systems with many similar, barely interacting parts are clearly quite
simple. Contrasting a pile of sand with a silicon chip, we naturally
want our complexity measure to capture the inherent complexity in the
silicon chip, even if they're made of similar numbers of parts of
similar material.

Since the pile of sand case indicates complexity is not simply the
number of components making up a system, the relationships between
components clearly contribute to the overall complexity. One can start
by caricaturing the system as a {\em graph} --- replacing the
components by abstract {\em vertices} or {\em nodes} and relationships
between nodes by abstract {\em edges} or {\em arcs}. 

Graph theory~\cite{Diestel05} was founded by Euler in the 18th century
to solve the famous K\"onigsberg bridge problem. However, until the
1950s, only simple graphs that could be analysed in toto were
considered. Erd\"os and Renyi~\cite{Erdos-Renyi59}
introduced the concept of a {\em random graph}, which allowed one to
  treat large complex graphs statistically. Graphs of various sorts
  were readily recognised in nature, from food webs, personal or
  business contacts, sexual relations and the Internet amongst others.
  However, it soon became apparent that natural networks often had
  different statistical properties than general random graphs. Watts
  and Strogatz~\cite{Watts-Strogatz98} introduced the {\em
    small world} model, which has sparked a flurry of activity in
  recent years to measure networks such as the Internet, networks of
  collaborations between scientific authors and food webs in
  ecosystems~\cite{Albert-Barabasi01}.

Graph theory provides a number of measures that can stand in for
complexity. A number of these are illustrated in Figure
\ref{graphTheoreticMeasures}. The simplest of these is the {\em
connectivity} of a graph, namely the number of edges connecting
vertices of the graph. A fully connected graph, however, is no more
complex than one that is completely unconnected. As connectivity
increases from zero, a {\em percolation threshold} is reached where
the graph changes from being mostly discontinuous to mostly
continuous. The most complex systems tend to lie close to the
percolation threshold. Another graph measure used is the {\em
cyclomatic number} of a graph, basically the number of independent
loops it contains. The justification for using cyclomatic number as a
measure of complexity is that feedback loops introduce nonlinearities
in the system's behaviour, that produce complex behaviour.

% diagrams of various graph theoretic concepts.
% add in some discussion of the history of graph theory.

\def\graphnodes#1{
\begin{pspicture}(3,2)
\cnodeput(1,0){A}{A}
\cnodeput(0,1){B}{B}
\cnodeput(1,2){C}{C}
\cnodeput(3,1.5){D}{D}
\cnodeput(2,.5){E}{E}
#1
\end{pspicture}
}

\begin{figure}
\begin{pspicture}(-.5,-2.5)(14.5,9)
%\psgrid

\rput[l](0,6)
{\psset{unit=2}
\graphnodes{
        \ncline{A}{B}
        \ncline{B}{C}
        \ncline{C}{D}
        \ncline{D}{E}
        \ncline{B}{D}
        \ncline{C}{E}
        }
}

\rput[l](7,6){
\begin{tabular}{c@{\hspace{1ex}=\hspace{1ex}}l}
nodes & 5\\
connectivity & 6/25\\
cyclomatic no. & 2\\
spanning trees & 4\\
height (depth) & 2\\
\end{tabular}
}

\rput(-3,-2){
\psframe[linestyle=dashed](5.5,-.5)(14.5,6)
\rput(10,3){{\bf Spanning Trees}}
\rput(7.5,4){
  \graphnodes{
    \ncline{A}{B}\ncline{B}{C}\ncline{C}{D}\ncline{D}{E}
    }
  }
\rput(12.5,4){
  \graphnodes{
    \ncline{A}{B}\ncline{B}{C}\ncline{B}{D}\ncline{D}{E}
    }
  }
\rput(7.5,1){
  \graphnodes{
    \ncline{A}{B}\ncline{B}{C}\ncline{C}{D}\ncline{C}{E}
    }
  }
\rput(12.5,1){
  \graphnodes{
    \ncline{A}{B}\ncline{B}{C}\ncline{B}{D}\ncline{C}{E}
    }
  }
}

\end{pspicture}
\caption{Various graph theoretic measures for a simple graph. The
spanning trees are shown in the dashed box}
\label{graphTheoreticMeasures}
\end{figure}

Related to the concept of cyclomatic number is the {\em number of
spanning trees} of the graph. A spanning tree is a subset of the graph
that visits all nodes but has no loops (ie is a tree). A graph made up
from several disconnected parts has no spanning tree. A tree has
exactly one spanning tree. The number of spanning trees increases
rapidly with the cyclomatic number.

The height of the flattest spanning tree, or equivalently the maximum
number of hops separating two nodes on the graph
%\footnote{popularised
%in the movie {\em six degrees of separation} --- which refers to the
%maximum number of acquaintances connecting any two people in the
%World} 
is another useful measure related to complexity, usually called
the {\em diameter}. Networks having small degrees of separation (so
called {\em small world networks}) tend to support more complex
dynamics than networks having a large degree of separation. The reason
is that any local disturbance is propagated a long way through a small
world network before dying out, giving rise to chaotic dynamics,
whereas in the other networks, disturbances remain local, leading to
simpler linear dynamics.

\section{Erdos-Renyi random graphs, small world networks, scale-free
networks}

When considering the statistical properties of large networks, it is
useful to randomly generate networks having particular properties
from simple models. These may be used, for instance, as null models,
to determine if the network being studied has attributes that are
statistically significantly different from the null model.

The simplest such random model was introduced in the 1950s by Erd\"os
and R\'enyi~\cite{Erdos-Renyi59}. Starting with $n$ nodes, add $\ell$
edges by randomly selecting pairs of nodes and attaching an
edge. Equivalently, one can add an edge between any pair of nodes with
probability $p=\ell/n(n-1)$. Erd\"os-R\'enyi graphs exhibit a Gaussian
degree distribution, and substantially more clustering compared with
graphs embedded in a low dimensional space (eg wireframe meshes).

Graphs embedded in a low dimensional space\footnote{the regular ones are
usually called Cartesian graphs, as the nodes are just the points
whose Cartesian coordinates are integral} have a high graph diameter
(many edges need to be traversed to pass from one node to another
randomly chosen node). By contrast, random graphs of sufficiently high
connectivity tend to have low diameter, between any randomly chosen
pair of nodes, there will be a path traversing only a few edges, a
property called {\em small world}. One can construct small world graphs
in between Cartesian graphs and random graphs by starting with a
Cartesian graph, and randomly rewiring a small proportion of the edges.

Many real world networks exhibit a {\em scale free} property, with the
node degree distribution following a power law. One popular algorithm
for generating these sorts of networks is {\em preferential
attachment}, which involves adding links preferentially to nodes with
higher degree in a ``rich gets richer'' effect~\cite{Barabasi-Albert99}.

\section{Graph entropy}

There is a long tradition of applying information theory to graph
structures, starting with Rashevsky~\cite{Rashevsky55},
Trucco~\cite{Trucco56} and
Mowshowitz~\cite{Mowshowitz68a,Mowshowitz68b,Mowshowitz68c,Mowshowitz68d}. A
recent, detailed review can be found in
~\cite{Dehmer-Mowshowitz11}. 

Given a graph $G=V\times E$ of nodes $V$ and links $E$, and {\em graph
invariant} function $\alpha$ defined on the nodes:
\begin{eqnarray*}
\alpha: V &\rightarrow& A\\
\end{eqnarray*}
we can form the graph entropy measure 
\begin{equation}\label{graphEntropy}
S(G,\alpha) = |V|\log |V| - \sum_{a\in A} |\alpha^{-1}(a)|\log|\alpha^{-1}(a)|,
\end{equation}
where $|\cdot|$ is the usual notation for set cardinality. 
The sum in eq (\ref{graphEntropy}) is over sets of nodes that are
equivalent under the map $\alpha$. This plays the analogous role to
the observer function $O(x)$ mentioned previously. In
\S\ref{IBCnetwork}, we will use the automorphism relation between
graphs as the observer function --- the corresponding $\alpha$
function maps nodes to their {\em orbits}. Other graph invariants
have also been used in the literature, such as node degree, or level in a
tree structure.

A very similar measure to (\ref{graphEntropy}) is obtained by
averaging the information contained in each orbit:
\begin{equation}
I(G,\alpha) = -\sum_{a\in A} P_i \log P_i = 
-\sum_{a\in A}
\frac{|\alpha^{-1}(a)|}{|V|}\log\frac{|\alpha^{-1}(a)|}{|V|}
\end{equation}

One can likewise form similar measures by considering graph invariants
over links, rather than nodes.

\section{Information based complexity of unweighted, unlabeled,
undirected networks}\label{IBCnetwork}

In order to compute the complexity according to equation
(\ref{complexity}), it is necessary to fix two things: a bitstring
representation ({\em description}) of the item in question, and a
means of determining if two descriptions describe the same object. 

In the case of graphs, we consider two graphs to be identical if and
only if a permutation of nodes exists that allows the nodes of one
graph to be placed in a 1--1 correspondence with the nodes of the
other. In other words, an {\em automorphism}. If either the nodes or
edges are labelled, or a dynamic process is defined on the network, a
situation we will consider in subsequent sections, then the labelling
(or process in that case) must also be preserved by the automorphism.

One very simple implementation language for undirected graphs is to
label the nodes $1\ldots n$, and the links by the pair $(i,j),\; i<j$ of
nodes that the links connect. The linklist can be represented simply
by an $L=n(n-1)/2$ length bitstring, where the $\frac12j(j-1)+i$th
position is 1 if link $(i,j)$ is present, and 0 otherwise.

The directed case requires doubling the size of the
linklist, ie or $L=n(n-1)$. We also need to prepend the string with
the value of $N$ in order to make it prefix-free --- the simplest
approach being to interpret the number of leading 1s as the number $n$,
which adds a term $n+1$ to the measured complexity.

This proposal was analysed in~\cite{Standish05a}, and has the
unsatisfactory property that the fully connected or empty networks are
maximally complex for a given node count. An alternative scheme is to
also include the link count as part of the prefix, and to use binary
coding for both the node and link counts~\cite{Standish10a}. The sequence will start with
$\lceil\log_2n\rceil$ 1's, followed by a zero stop bit, so the prefix will be
$2\lceil\log_2n\rceil+\lceil\log_2L\rceil+1$ bits.

This scheme entails that some of bitstrings are not valid networks,
namely ones where the link count does not match the number of 1s in
the linklist. We can, however, use rank encoding~\cite{Myrvold-Ruskey01} of the
linklist to represent the link pattern. The number of possible
linklists corresponding to a given node/link specification is given by
\begin{equation}
\Omega = \left(\begin{array}{c} 
L \\
l \\
\end{array}
\right) = \frac{L!}{(L-l)! l!}
\end{equation}
This will have a minimum value of 1 at $l=0$ (empty network) and
$l=L$, the fully connected network.

Finally, we need to compute $\omega$ of the linklist, which is just
the total number of possible renumberings of the nodes ($n!$), divided
by the size of the graph automorphism group $|{\cal A}|$, which can be
practically computed by Nauty~\cite{McKay81}, or a number of other
algorithms which exhibit better performance on sparsely linked
networks~\cite{Darga-etal08, Junttila-Kaski07,cordella-eta01}. With
$\omega$ computed, the complexity ${\cal C}$ of the network is given
by (\ref{complexity}).

\begin{figure}
\begin{center}
\resizebox{\textwidth}{!}{\includegraphics{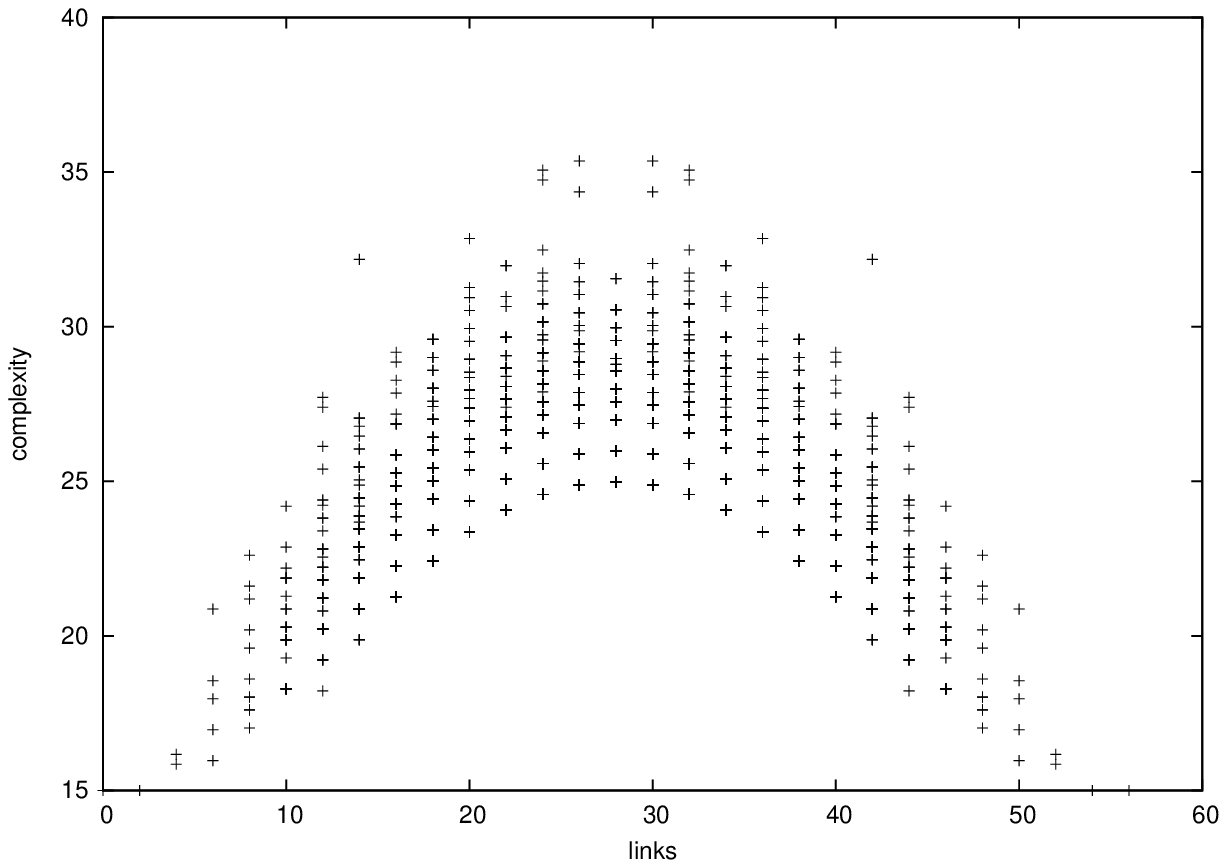}}
\end{center}
\caption{The new complexity measure as a function of link count for
  all networks with 8 nodes. This shows the strong dependence of
  complexity on link count, and the symmetry between networks and
  their complements.
}
\label{l-C}
\end{figure}

A network $A$ that has a link wherever $B$ doesn't, and vice-versa
might be called a complement of $B$. A bitstring for $A$ can be found
by inverting the 1s and 0s in the linklist part of the network
description. Obviously, $\omega(A,L)=\omega(B,L)$, so the complexity
of a network is equal to that of its complement, as can be seen in
Figure \ref{l-C}.

A connection between ${\cal C}$ and the graph entropy $S$ defined in
equation (\ref{graphEntropy}) can be made by noting that the size of
the automorphism group is simply the product of the sizes
of the orbits:
\begin{equation}
|{\cal A}| = \prod_{a\in A} |\alpha^{-1}(a)|!
\end{equation}
Using the Stirling approximation ($\log x!\approx x\log x$), we may
write
\begin{eqnarray}
\omega &=& \frac {n!}{|{\cal A}|} = \frac{|V|!}{\prod_{a\in A}
|\alpha^{-1}(a)|!} \nonumber\\
\log\omega &\approx& |V|\log|V| - \sum_{a\in
A}|\alpha^{-1}(a)|\log|\alpha^{-1}(a)| = S
\end{eqnarray}

\section{Motif expansion}

Adami et al.~\cite{Adami-etal11} introduce the concept of {\em motif
entropy}. By breaking the network into motifs (eg pairs of nodes
connected by a link, triangles, quads, 3-pointed stars, etc), and
forming the Shannon entropy $H=-\sum_ip_i\log_2p_i$, where $p_i$ are
the probabilities of the various motifs occurring, one gets a measure
which they call motif entropy. This should converge to
(\ref{complexity}) as more motifs are included, in just the same way
as (\ref{shannon entropy}) converges to (\ref{shannon}) as longer
sequences are included. Adami et al. restrict themselves to motifs of
two and three nodes only in examining the neural network of {\em
C. elegans}, and also in examining the epistatic interaction networks
in the Avida digital organism system. They show that this suffices to
capture meaningful adaptive information about the systems, but not
whether it captures all the pertinent information. Further work
linking motif entropy with network complexity is called for.

\section{Labelled networks}

If all nodes are labelled with distinct labels, the network is
uniquely specified by the node and link counts, along with a
rank-encoded linklist. In this case, eq (\ref{complexity}) can be expressed
analytically: 
\begin{equation}\label{distinct label} 
{\cal C} =
2\lceil\log_2n\rceil+\lceil\log_2L\rceil+1 + \left\lceil\log_2
\frac{L!}{(L-l)! l!}\right\rceil 
\end{equation} 
% Is this formula given by Bouton & Wallace '69?

In the case where the labels are not distinct, the network is often
said to be {\em coloured}\footnote{Eg. if there are three distinct
labels, they may as well be red, green and blue.}. Not much work has
been done calculating the complexity of coloured networks, but
recently Adami et al.~\cite{Adami-etal11} tackled the problem. In that
paper, motif expansion was used to approximate the complexity.

One can use eq (\ref{complexity}) directly, provided one had an
algorithm for computing the size of the automorphism group that leaves
the colour labels invariant. This is still an open problem, but in
principle, existing automorphism algorithms should be able to be
adapted.

However, special cases exist where the coloured network complexity
reduces to uncoloured network complexity. For example, if all nodes
within a  colour grouping have distinct degree, then the problem is identical to
the distinct label case, and equation (\ref{distinct label}) can be
used. Similarly, if all nodes of the same degree have the same colour,
then the coloured network complexity is identical to that of the
uncoloured network. 

\section{Weighted networks}

Whilst the information contained in link weights might be significant
in some circumstances (for instance the weights of a neural network
can only be varied in a limited range without changing the overall
qualitative behaviour of the network), of particular theoretical
interest is to consider the weights as continuous parameters
connecting one network structure with another. For instance if a
network $X$ has the same network structure as the unweighted graph A,
with $b$ links of weight 1 describing the graph $B$ and the
remaining $a-b$ links of weight $w$, then we would like the network
complexity of $X$ to vary smoothly between that of $A$ and $B$ as $w$
varies from 1 to 0.~\cite{Gornerup-Crutchfield08} introduced a similar
measure.

The most obvious way of defining this continuous complexity measure is
to start with normalised weights $\sum_iw_i=1$. Then arrange the links
in weight order, and compute the complexity of networks with just
those links of weights less than $w$. The final complexity value is
obtained by integrating:
\begin{equation}\label{weighted C}
{\cal C}(X=N\times L) = \int_0^1 {\cal C}(N\times\{i\in L: w_i<w\}) dw
\end{equation}
Obviously, since the integrand is a stepped function, this is computed
in practice by a sum of complexities of partial networks.

\section{Empirical results of real network data, and artificially generated
   networks}

Table \ref{Tab1} shows the complexities of a number of well-known real
world networks~\cite{Standish10a}. Also shown is the average complexity of 1000 shuffled
networks. Shuffling the links of a network produces an Erd\"os-R\'enyi
random network with an identical link weight distribution to the
original network.

In most cases, there is a statistically siginificant difference
between the real network complexity and the shuffled version,
indicating that the network structure encodes siginificant
information.

In~\cite{Standish10a}, several evolutionary systems from Artificial
Life are also analysed in the same way, as well as networks generated
by the Erd\"os-R\'enyi process and the Barab\'asi-Albert {\em
preferential attachment process}~\cite{Barabasi-Albert99}. Networks
derived from evolutionary process exhibited the same sort of
complexity excess as the real world network, but networks created from
purely random processes did not, indicating the information hoarding
nature of adaption.

%\singlespace
\begin{table}
\begin{tabular}{lrrrrrr}
\small
Dataset & nodes & links &${\cal C}$ & $e^{\langle\ln{\cal C}_\mathrm{ER}\rangle}$ & ${\cal
  C}-e^{\langle\ln{\cal C}_\mathrm{ER}\rangle}$ & $\frac{|\ln{\cal
  C}-\langle\ln{\cal C}_\mathrm{ER}\rangle|}{\sigma_\mathrm{ER}}$ \\\hline
celegansneural & 297 & 2345 & 442.7 &251.6 &191.1 &29\\
lesmis & 77 & 508 &199.7 &114.2 &85.4 &24\\
adjnoun & 112 & 850 & 3891 & 3890 & 0.98 & $\infty$\\
yeast & 2112 & 4406 & 33500.6 & 30218.2 & 3282.4 & 113.0\\
celegansmetabolic & 453 & 4050 & 25421.8 & 25387.2 & 34.6&$\infty$  \\
baydry & 128 & 2138 &126.6 &54.2 &72.3 &22\\
baywet & 128 & 2107 &128.3 &51.0 &77.3 &20\\
cypdry & 71 & 641 & 85.7 &44.1 &41.5 &13\\
cypwet & 71 & 632 & 87.4 &42.3 &45.0 &14\\
gramdry & 69 & 911 & 47.4 &31.6 &15.8 &10\\
gramwet & 69 & 912 &54.5 &32.7 &21.8 &12\\
Chesapeake& 39& 177& 66.8& 45.7& 21.1& 10.4\\
ChesLower& 37& 178& 82.1& 62.5& 19.6& 10.6\\
ChesMiddle& 37& 208& 65.2& 48.0& 17.3& 9.3\\
ChesUpper& 37& 215& 81.8& 60.7& 21.1& 10.2\\
CrystalC& 24& 126& 31.1& 24.2& 6.9& 6.4\\
CrystalD& 24& 100& 31.3& 24.2& 7.0& 6.2\\
Everglades &69& 912& 54.5& 32.7& 21.8& 11.8\\
Florida& 128& 2107& 128.4& 51.0& 77.3& 20.1\\
Maspalomas& 24& 83& 70.3& 61.7& 8.6& 5.3\\
Michigan& 39& 219& 47.6& 33.7& 14.0& 9.5\\
Mondego& 46& 393& 45.2& 32.2& 13.0& 10.0\\
Narragan& 35& 219& 58.2& 39.6& 18.6& 11.0\\
Rhode& 19& 54& 36.3& 30.3& 6.0& 5.3\\
StMarks& 54& 354& 110.8& 73.6& 37.2& 16.0\\
PA1 & 100& 99 & 98.9& 85.4 & 13.5& 2.5\\
PA3 & 100& 177 & 225.9 & 207.3& 18.6& 3.0
\end{tabular}
\caption{
  Complexity values of several freely available network
  datasets. celegansneural, lesmis and adjnoun are available from Mark
  Newman's website, % (http://www-personal.umich.edu/~mejn/netdata/),
  representing the neural network of the {\em C. elegans}
  nematode~\cite{Watts-Strogatz98}, the coappearance of characters in
  the novel {\em Les Mis\'erables} by Victor Hugo~\cite{Knuth93} and
  the adjacency network of common adjectives and nouns in the novel {\em David
  Copperfield} by Charles Dickens~\cite{Newman06}. The metabolic data of
  {\em C. elegans}~\cite{Duch-Arenas05} and protein interaction network
  in yeast~\cite{Jeong-etal01} are available from Duncan Watt's
  website. PA1 and PA3 are networks generated via preferential
  attachment with in degree of one or three respectively, and uniformly
  distributed link weights. The other datasets are food webs 
  available from the Pajek website
 ~\cite{Christian-Luczkovich99,Ulanowicz-etal98,Ulanowicz-etal00,Hagy02,Baird-etal98,Almunia-etal99,Baird-Ulanowicz89}. For
  each network, the number of 
  nodes and links are given, along with the computed complexity ${\cal
  C}$. In the fourth column, the original network is shuffled 1000
  times, and the logarithm of the complexity is averaged
  ($\langle\ln{\cal C}_\mathrm{ER}\rangle$). The fifth
  column gives the difference between these two values, which
  represents the information content of the specific arrangement of
  links. The final column gives a measure of the significance of this
  difference in terms of the number of standard deviations (``sigmas'') of the
  distribution of shuffled networks. In two examples, the
  distributions of shuffled networks had zero standard deviation, so
  $\infty$ appears in this column.
}
\label{Tab1}
\end{table}

\section{Extension to processes on networks}

What has been discussed up until now is the static, or {\em
structural} complexity of a network. Often, a dynamic process occurs
on a network, such as neural network dynamics, or the ecological
dynamics of a foodweb. One is interested in the amount of complexity
contributed to the process by the network structure. Since two
distinct networks with the same attached dynamics may well be
considered identical in some context (perhaps by having the same
attractor basins, for instance), then in general the {\em dynamic
complexity} is less than the {\em structural complexity}.

With continuous processes, there is a practical difficulty of
establishing whether two networks generate the same process,
particularly if there is an element of stocasticity involved. When
comparing continuous-valued time series, one would need to choose a
metric over (in general) a multidimensional space, and an error
threshold within which two time series are considered the
same. Furthermore, a maximum time period for comparison needs to be
chosen, as dynamical chaos effects are likely to render two
arbitrarily close trajectories significantly different after a finite
period of time.

If the initial transients of the processes aren't important, one could
compare basins of attraction instead, which only eliminates the choice
of time period in the comparison.

With discrete (or symbolic) processes, the problem is conceptually
simpler in that one can determine if two networks generate identical
processes according to an observer function $O(x)$.

Nevertheless, the computational complexity of this approach rules it out
for all but the simplest of networks.

An alternative approach is given by considering the amount of
information flowing between nodes, a notion known as {\em transfer entropy}.

\section{Transfer Entropy}

Given a time series $X_t$, let $X^-_t=\{X_t, X_{t-1}, X_{t-2}, \ldots\}$
be the {\em history} of $X$ up to time $t$.

The mutual information $I(X_{t+1}: Y^-_t)$ gives a measure of the extent
to which the history of $Y$ disambiguates the future of $X$. However,
$Y$ may itself depend on the past of $X$, giving rise to spurious
directional effects between $Y$ and $X$~\cite{Kaiser-Scheiber02}. So
we should also condition on the past of $X$, giving rise to the notion
of {\em transfer entropy}:
\begin{equation}\label{transfer entropy}
{\cal T}_{Y\rightarrow X} = I(X_{t+1}: Y^-_t|X^-_t)
\end{equation}

Transfer entropy has been applied to random boolean
networks~\cite{Lizier-etal11}, but is more usually used to infer
network structure from time series data such as neural networks
~\cite{Wibral-etal11} or genetic regulatory networks
~\cite{Tung-etal07}.

We may also condition the transfer entropy on the state of the rest of
system ${\bf U}$ (not including $X$ or $Y$):

\begin{equation}\label{complete transfer entropy}
{\cal T}_{Y\rightarrow X|{\bf U}} = I(X_{t+1}: Y^-_t|X^-_t, {\bf U}^-_t).
\end{equation}
Lizier et al~\cite{Lizier-etal11} call (\ref{transfer entropy}) the
{\em apparent} transfer entropy, and (\ref{complete transfer entropy})
the {\em complete} transfer entropy. They find that apparent transfer
entropy is maximised around a critical point corresponding to a
connectivity of around 2 links per node, whereas complete transfer
entropy rises near the critical point, and continues to rise as
connectivity increases and the system moves into the chaotic regime,
up to a connectivity of 5 links per node.

A related concept to transfer entropy is {\em Granger
causality}. Granger causality between two nodes $X$ and $Y$ is found
by considering a linear multivariate model of the lags
\begin{displaymath}
X_{t+1} = \sum_k A_kX_{t-k} + \sum_k B_kY_{t-k} + \epsilon
\end{displaymath}
and a restricted linear model with the $Y$ terms removed:
\begin{displaymath}
X_{t+1} = \sum_k C_kX_{t-k} + \epsilon'.
\end{displaymath}
If the former full model gives a statistically significant better fit
to the data than the latter restricted model, we say that $Y$ {\em
Granger causes} $X$. To quantify the statistical significance, we use
the F-statistic
\begin{equation}
{\cal F}_{Y\rightarrow X} = \ln
\frac{\langle\epsilon'^2\rangle}{\langle\epsilon^2\rangle}
\end{equation}
as the variance of residuals in the restricted model
($\langle\epsilon'^2\rangle$) will be more than that of the full model. 

Granger causality has the advantages of being computationally simpler,
as well as having an interpretation in terms of statistical
significance. The downside is that it captures linear relationships
only, whereas transfer entropy is model-free, capturing all that is
relevant between entities. The two concepts are very closely related,
and for the special case of Gaussian processes, are identical up to a
factor of 2~\cite{Barnett-etal09}.

The models used in Granger causality may also include the remainder of
the system ${\bf U}$, and this is used for computing the {\em causal
density} of the system, which is the proportion of pairs of nodes
where one node {\em Granger causes} the other~\cite{Seth-etal11}.  The
measure has a minimum for weakly interacting nodes, and likewise for
strongly interacting nodes (as everything influences everything else,
so is conditioned out). It has a maximum in between, expressing a
balance between integration and segregation in a system. It is very
similar to an earlier measure proposed by Tononi, Sporns and Edelman
(TSE complexity)~\cite{Tononi-etal94} which is based on mutual
information across bipartitions of the network rather than transfer
entropy.

Like ${\cal C}$ from (\ref{weighted C}), both TSE complexity and
causal density are minimal for sparse and dense networks, rising to a
maximum value in between. However the maximum value of causal density
occurs around the order-chaos transition (approx 2 links per node),
which is a distinctly different peak to that of structural complexity,
which is at a maximum at $n/2$ links per node.

\section{Medium Articulation}

Wilhelm~\cite{Wilhelm-Hollunder07, Kim-Wilhelm08} introduced a new
complexity like measure that addresses the intuition that complexity
should be minimal for the empty and full networks, and peak for
intermediate values (like figure \ref{l-C}). It is obtained by
multiplying the mutual information between all pairs of nodes by the
conditional entropy across all links (which they call the {\em
redundancy}). The resulting measure also has a quality of measuring
the segregation/integration balance reminiscent of causal density.

Precisely, medium articulation is given by
\begin{equation}
\textrm{MA} = -\sum_{ij} w_{ij} \log \frac{w_{ij}}{\sum_k w_{ik}\sum_k
w_{kj}}
        \times \sum_{ij} w_{ij} \log \frac{w_{ij}^2}{\sum_k w_{ik}\sum_k
w_{kj}},
\end{equation}
where $w_{ij}$ is the normalised weight ($\sum_{ij}w_{ij}=1$) 
of the link from node $i$ to node $j$. It should be noted that this is
just the product of the two terms $A$ and $\Phi$ representing the
degree of constraint and the extent of freedom of the system in
Ulanowicz's paper in this volume~\cite{Ulanowicz12}.
 
\begin{figure}
\input{cma}
\caption{Medium Articulation plotted against complexity for 1000
randomly sampled Erd\"os-R\'enyi graphs up to order 500.}
\label{CMA}
\end{figure}

Figure \ref{CMA} shows medium articulation plotted against ${\cal
C}$ for a sample of 1000 Erd\"os-R\'enyi networks up to order
500. There is no clear relationship between medium articulation and
complexity for the average network. Medium articulation does not
appear to discriminate between complex networks. however if we restrict our
attention to simple networks (Figures \ref{CMA1} and \ref{CMA2})
medium articulation is strongly correlated with complexity, and so
could be used as a proxy for complexity for these cases.

This lends some credence to the notion that causal density, TSE
complexity and network complexity are all related.

\begin{figure}
\input{cma1}
\caption{Medium Articulation plotted against complexity for 1000
randomly sampled Erd\"os-R\'enyi graphs up to order 500 with no more
than $2n$ links.}
\label{CMA1}
\end{figure}

\begin{figure}
\input{cma2}
\caption{Medium Articulation plotted against complexity for 1000
randomly sampled Erd\"os-R\'enyi graphs up to order 500 with more
than $n(n-5)/2$ links.}
\label{CMA2}
\end{figure}

\section{Conclusion}

In this chapter, a number of information-based measures of network
complexity are considered. Measures of structural complexity are found
to be related to each other, and similarly information flow measures
of dynamic complexity are also found to be related. It would seem
plausible that dynamic complexity measures should be related to
structural complexity when the dynamical processes are in some sense
generic, or uncoloured, but at this stage, such a conjecture remains
unproven. For relatively simple processes such as Gaussian processes,
and the Random Boolean Networks studied by Lizier et al., the
behaviour of a dynamical complexity measure has a peak at much lower
connectivities than the peak exhibited by the structural complexity
measure. More work is required to clarify the relationship between
dynamical and strucural complexity of networks.

\bibliographystyle{plain}
\bibliography{rus}

\begin{thebibliography}{10}

\bibitem{Adami-etal11}
C.~Adami, J.~Qian, M.~Rupp, and A.~Hintze.
\newblock Information content of colored motifs in complex networks.
\newblock {\em Artificial Life}, 17:375--390, 2011.

\bibitem{Adami98a}
Chris Adami.
\newblock {\em Introduction to Artificial Life}.
\newblock Springer, 1998.

\bibitem{Albert-Barabasi01}
R\'eka Albert and Albert-L\'aszl\'o Barab\'asi.
\newblock Statistical mechanics of complex networks.
\newblock {\em Reviews of Modern Physics}, 74:47, 2002.

\bibitem{Almunia-etal99}
J.~Almunia, G.~Basterretxea, J.~Aristegui, and R.E. Ulanowicz.
\newblock Benthic- pelagic switching in a coastal subtropical lagoon.
\newblock {\em Estuarine, Coastal and Shelf Science}, 49:363--384, 1999.

\bibitem{Baird-etal98}
D.~Baird, J.~Luczkovich, and R.~R. Christian.
\newblock Assessment of spatial and temporal variability in ecosystem
  attributes of the {St Marks National Wildlife Refuge}, {Apalachee Bay,
  Florida}.
\newblock {\em Estuarine, Coastal, and Shelf Science}, 47:329--349, 1998.

\bibitem{Baird-Ulanowicz89}
D.~Baird and R.E. Ulanowicz.
\newblock The seasonal dynamics of the {Chesapeake Bay} ecosystem.
\newblock {\em Ecological Monographs}, 59:329--364, 1989.

\bibitem{Barabasi-Albert99}
Albert-L\'asl\'o Barab\'asi and R\'eka Albert.
\newblock Emergence of scaling in random networks.
\newblock {\em Science}, 286:509--512, 1999.

\bibitem{Barnett-etal09}
L.~Barnett, A.B. Barrett, and A.K. Seth.
\newblock Granger causality and transfer entropy are equivalent for gaussian
  variables.
\newblock {\em Physical review letters}, 103(23):238701, 2009.

\bibitem{Christian-Luczkovich99}
R.R. Christian and J.J. Luczkovich.
\newblock Organizing and understanding a winter's seagrass foodweb network
  through effective trophic levels.
\newblock {\em Ecological Modelling}, 117:99--124, 1999.

\bibitem{cordella-eta01}
L.P. Cordella, P.~Foggia, C.~Sansone, and M.~Vento.
\newblock An improved algorithm for matching large graphs.
\newblock In {\em 3rd IAPR-TC15 workshop on graph-based representations in
  pattern recognition}, pages 149--159, 2001.

\bibitem{Darga-etal08}
P.~T. Darga, K.~A. Sakallah, and I.~L. Markov.
\newblock Faster symmetry discovery using sparsity of symmetries.
\newblock In {\em Proceedings of the 45st Design Automation Conference},
  Anaheim, California, June 2008.

\bibitem{Dehmer-Mowshowitz11}
M.~Dehmer and A.~Mowshowitz.
\newblock A history of graph entropy measures.
\newblock {\em Information Sciences}, 181(1):57--78, 2011.

\bibitem{Diestel05}
Reinhard Diestel.
\newblock {\em Graph Theory}.
\newblock Springer, Berlin, 3rd edition, 2005.

\bibitem{Duch-Arenas05}
J.~Duch and A.~Arenas.
\newblock Community identification using extremal optimization.
\newblock {\em Physical Review E}, 72:027104, 2005.

\bibitem{Erdos-Renyi59}
Paul Erd\"os and Alfr\'ed R\'enyi.
\newblock On random graphs.
\newblock {\em Publ. Math. Dubrecen}, 6:290--291, 1959.

\bibitem{Gell-Mann94}
Murray Gell-Mann.
\newblock {\em The Quark and the Jaguar: Adventures in the Simple and the
  Complex}.
\newblock Freeman, 1994.

\bibitem{Gornerup-Crutchfield08}
Olof G\"ornerup and James~P. Crutchfield.
\newblock Hierarchical self-organization in the finitary process soup.
\newblock {\em Artificial Life}, 14:245--254, 2008.

\bibitem{Hagy02}
J.D. Hagy.
\newblock {\em Eutrophication, hypoxia and trophic transfer efficiency in
  {Chesapeake Bay}}.
\newblock PhD thesis, University of Maryland at College Park (USA), 2002.

\bibitem{Jeong-etal01}
Hawoong Jeong, Sean Mason, Albert-L\'aszl\'o Barab\'asi, and Zolt\'an~N.
  Oltvai.
\newblock Centrality and lethality of protein networks.
\newblock {\em Nature}, 411:41, 2001.

\bibitem{Junttila-Kaski07}
Tommi Junttila and Petteri Kaski.
\newblock Engineering an efficient canonical labeling tool for large and sparse
  graphs.
\newblock In {\em Proceedings of the Ninth Workshop on Algorithm Engineering
  and Experiments (ALENEX07)}, pages 135--149. SIAM, 2007.

\bibitem{Kaiser-Scheiber02}
A.~Kaiser and T.~Schreiber.
\newblock Information transfer in continuous processes.
\newblock {\em Physica D: Nonlinear Phenomena}, 166:43--62, 2002.

\bibitem{Kim-Wilhelm08}
Jongwang Kim and Thomas Wilhelm.
\newblock What is a complex graph.
\newblock {\em Physica A}, 387:2637--2652, 2008.

\bibitem{Knuth93}
D.~E. Knuth.
\newblock {\em The {Stanford GraphBase}: A Platform for Combinatorial
  Computing}.
\newblock Addison-Wesley, Reading, MA, 1993.

\bibitem{Li-Vitanyi97}
Ming Li and Paul Vit\'anyi.
\newblock {\em An Introduction to {Kolmogorov} Complexity and its
  Applications}.
\newblock Springer, New York, 2nd edition, 1997.

\bibitem{Lizier-etal11}
Joseph Lizier, Siddharth Pritam, and Mikhail Prokopenko.
\newblock Information dynamics in small-world boolean networks.
\newblock {\em Artificial Life}, 17:293--314, 2011.

\bibitem{McKay81}
Brendan~D. McKay.
\newblock Practical graph isomorphism.
\newblock {\em Congressus Numerantium}, 30:45--87, 1981.

\bibitem{Mowshowitz68a}
A.~Mowshowitz.
\newblock Entropy and the complexity of graphs: I. an index of the relative
  complexity of a graph.
\newblock {\em Bulletin of Mathematical Biology}, 30(1):175--204, 1968.

\bibitem{Mowshowitz68b}
A.~Mowshowitz.
\newblock Entropy and the complexity of graphs: Ii. the information content of
  digraphs and infinite graphs.
\newblock {\em Bulletin of Mathematical Biology}, 30(2):225--240, 1968.

\bibitem{Mowshowitz68c}
A.~Mowshowitz.
\newblock Entropy and the complexity of graphs: Iii. graphs with prescribed
  information content.
\newblock {\em Bulletin of Mathematical Biology}, 30(3):387--414, 1968.

\bibitem{Mowshowitz68d}
A.~Mowshowitz.
\newblock Entropy and the complexity of graphs: Iv. entropy measures and
  graphical structure.
\newblock {\em Bulletin of Mathematical Biology}, 30(4):533--546, 1968.

\bibitem{Myrvold-Ruskey01}
Wendy Myrvold and Frank Ruskey.
\newblock Ranking and unranking permutations in linear time.
\newblock {\em Information Processing Letters}, 79:281--284, 2001.

\bibitem{Newman06}
M.~E.~J. Newman.
\newblock Finding community structure in networks using the eigenvectors of
  matrices.
\newblock {\em Phys. Rev. E}, 74:036104, 2006.

\bibitem{Rashevsky55}
N.~Rashevsky.
\newblock Life, information theory, and topology.
\newblock {\em Bulletin of Mathematical Biology}, 17(3):229--235, 1955.

\bibitem{Seth-etal11}
Anil~K. Seth, Adam~B. Barrett, and Lionel Barnett.
\newblock Causal density and integrated information as measures of conscious
  level.
\newblock {\em Phil. Trans. R. Soc. A}, 369:3748--3767, 2011.

\bibitem{Shannon49}
Claude~E. Shannon.
\newblock {\em The Mathematical Theory of Communication}.
\newblock U. Illinois Press, Urbana-Champaign, IL, 1949.

\bibitem{Standish01a}
Russell~K. Standish.
\newblock On complexity and emergence.
\newblock {\em Complexity International}, 9, 2001.
\newblock arXiv:nlin.AO/0101006.

\bibitem{Standish05a}
Russell~K. Standish.
\newblock Complexity of networks.
\newblock In Abbass et~al., editors, {\em Recent Advances in Artificial Life},
  volume~3 of {\em Advances in Natural Computation}, pages 253--263, Singapore,
  2005. World Scientific.
\newblock arXiv:cs.IT/0508075.

\bibitem{Standish10a}
Russell~K. Standish.
\newblock Complexity of networks (reprise).
\newblock {\em Complexity}, 17:50--61, 2012.
\newblock arXiv: 0911.348.

\bibitem{Tononi-etal94}
G.~Tononi, O.~Sporns, and G.~M. Edelman.
\newblock A measure for brain complexity: relating functional segregation and
  integration in the nervous system.
\newblock {\em Proc. Natl Acad. Sci. USA}, 91:5033--5037, 1994.

\bibitem{Trucco56}
E.~Trucco.
\newblock A note on the information content of graphs.
\newblock {\em Bulletin of Mathematical Biology}, 18(2):129--135, 1956.

\bibitem{Tung-etal07}
Thai~Quang Tung, Taewoo Ryu, K.H. Lee, and Doheon Lee.
\newblock Inferring gene regulatory networks from microarray time series data
  using transfer entropy.
\newblock In {\em Twentieth IEEE International Symposium on Computer-Based
  Medical Systems}, pages 383--388, 2007.

\bibitem{Ulanowicz-etal98}
R.E. Ulanowicz, C.~Bondavalli, and M.S. Egnotovich.
\newblock Network analysis of trophic dynamics in {South Florida} ecosystem,
  {FY} 97: The {Florida Bay} ecosystem.
\newblock Technical report, Chesapeake Biological Laboratory, Solomons, MD
  20688-0038 USA, 1998.
\newblock [UMCES]CBL 98-123.

\bibitem{Ulanowicz-etal00}
R.E. Ulanowicz, J.J. Heymans, and M.S. Egnotovich.
\newblock Network analysis of trophic dynamics in {South Florida} ecosystems,
  {FY} 99: The graminoid ecosystem.
\newblock Technical report, Chesapeake Biological Laboratory, Solomons, MD
  20688-0038 USA, 2000.
\newblock [UMCES] CBL 00-0176.

\bibitem{Ulanowicz12}
Robert Ulanowicz.
\newblock Circumscribed complexity in ecological networks.
\newblock In Abbe Mowshowitz and Matthias Dehmer, editors, {\em Advances in
  Network Complexity}. Wiley, 2012.

\bibitem{Watts-Strogatz98}
D.~J. Watts and S.~H. Strogatz.
\newblock Collective dynamics of `small-world' networks.
\newblock {\em Nature}, 393(6684):409--10, 1998.

\bibitem{Wibral-etal11}
M~Wibral, B~Rahm, M~Rieder, M~Lindner, R~Vicente, and J~Kaiser.
\newblock Transfer entropy in magnetoencephalographic data: quantifying
  information flow in cortical and cerebellar networks.
\newblock {\em Prog Biophys Mol Biol.}, 105:80--97, 2011.

\bibitem{Wilhelm-Hollunder07}
Thomas Wilhelm and Jens Hollunder.
\newblock Information theoretic description of networks.
\newblock {\em Physica A}, 385:385--396, 2007.

\end{thebibliography}
\end{document}